\documentclass[runningheads]{llncs}
\usepackage[T1]{fontenc}
\usepackage{graphicx,verbatim}
\usepackage{amsmath}
\usepackage{amssymb}
\usepackage{booktabs}
\usepackage{multirow}
\usepackage{colortbl}
\usepackage{graphicx}
\usepackage[table]{xcolor}
\usepackage{makecell} 

\usepackage{hyperref}
\begin{document}
\title{\textbf{E-MRL}: Cross-view Aligned Evidence-driven Multimodal Reinforcement Learning for Reliable 3D Tumor Analysis}

\author{Sijing Li\inst{1,2} \and
Zhongwei Qiu\inst{2,3}$^*$ \and  
Zhuoya Wang\inst{4} \and
Boxiang Yun\inst{5} \and
Zhenyu Yi\inst{6} \and
Jianwei Xu\inst{2} \and
Wenqiao Zhang\inst{1}$^*$ \and  
Yingda Xia\inst{2} \and
Ling Zhang\inst{2}
}


\authorrunning{S. Li et al.}
\institute{
Zhejiang University \and
DAMO Academy, Alibaba Group \and
Hupan Lab \and
Huazhong University of Science and Technology \and
East China Normal University \and
Shanghai Jiao Tong University
}
\titlerunning{Evidence-MRL for 3D Tumor Analysis}

\maketitle              
\begingroup
\renewcommand\thefootnote{}
\footnote{$^*$ Correspondence to: Wenqiao Zhang  (\href{wenqiaozhang@zju.edu.cn} {\texttt{wenqiaozhang@zju.edu.cn}}) and Zhongwei Qiu (\href{qiuzhongwei.qzw@alibaba-inc.com} {\texttt{qiuzhongwei.qzw@alibaba-inc.com}}).}
\addtocounter{footnote}{-1}
\endgroup

\begin{abstract}
While Vision-Language Models (VLMs) show great promise in volumetric medical report generation, they frequently suffer from visual hallucinations and a lack of grounding in 3D CT data. Current Supervised Fine-Tuning (SFT) and Reinforcement Learning (RL) strategies typically optimize text fidelity alone, essentially rewarding correct diagnoses derived from language priors rather than genuine visual perception. To address this, we propose cross-view aligned \textbf{\underline{E}}vidence-driven \textbf{\underline{M}}ultimodal \textbf{\underline{R}}einforcement \textbf{\underline{L}}earning (Evidence-MRL, noted as \textbf{E-MRL}), a reliable RL reasoning framework that formulates the generation process as a Markov Decision Process of ``diagnosis-localization-verification''. Unlike standard approaches, our model is explicitly trained to identify a ``key evidence slice'' alongside the global diagnostic report, grounding its findings in verifiable visual evidence. Crucially, we introduce a novel cross-view consistency reward, which validates the semantic alignment between the golden-standard report and a local visual re-query of the selected key slice, providing additional rewards for correctly-localized reasoning. Experiments on large-scale 3D CT tumor datasets demonstrate that E-MRL significantly reduces hallucinations and improves diagnostic accuracy compared to SFT and RL baselines, offering a clinically interpretable solution for visually-grounded and tumor analysis.


\end{abstract}

\section{Introduction}
The rapid advancement of Vision-Language Models (VLMs) has opened new frontiers in automated medical reasoning tasks~\cite{wu2025towards,sellergren2025medgemma,liu2026diyhealth,tu2024towards}, transitioning from 2D image analysis to complex 3D volumetric interpretation (e.g., CT and MRI scans)~\cite{bai2024m3d,blankemeier2024merlin,lin2026omnict}. 
However, the application of VLMs to 3D medical imaging is severely compromised by ``visual hallucinations''~\cite{jiangmedvr}, a phenomenon in which models generate clinically plausible but factually incorrect reports. Due to the excessive length of 3D sequences and the sparsity of pathological features, models often struggle to maintain attention on critical visual cues. 
Consequently, they tend to rely on language priors rather than genuine visual perception, fabricating lesion details that are inconsistent with or entirely absent from the input volume. Such ungrounded reasoning poses significant risks in clinical diagnostics. 

Existing alignment strategies, such as Supervised Fine-Tuning (SFT) and text-based Reinforcement Learning from Human Feedback (RLHF)~\cite{yu2024rlhf,litumorchain}, primarily optimize the linguistic coherence and factual correctness of reports at text level. However, these approaches share a fundamental limitation: they provide no process-level visual supervision over where the model looks during the reasoning process. If a model draws correct conclusions based on irrelevant slices (e.g., misidentifying a vessel cross-section as a tumor), standard loss functions fail to penalize this visual grounding error. The core challenge, therefore, is not merely improving textual quality, but establishing a verifiable correspondence between generated report content and its visual evidence in the 3D volume.

To address this challenge, inspired by the radiologist's diagnostic workflow of ``global scanning followed by local details verification'', we propose E-MRL, the first cross-view aligned \textbf{\underline{E}}vidence-driven \textbf{\underline{M}}ultimodal \textbf{\underline{R}}einforcement \textbf{\underline{L}}earning framework for reliable tumor analysis in 3D CT scans. Unlike conventional methods that rely solely on textual rewards, E-MRL frames report generation as a “diagnosis-localization-verification” Markov Decision Process and introduces multi-granularity textual and visual evidence-driven rewards based on a novel cross-view (global and local) consistency mechanism. While pioneers like RadGPT~\cite{bassi2025radgpt} rely on complex multi-stage, segmentation-assisted pipelines to surpass standard medical VLMs which often lack sufficient visual evidence, E-MRL represents the first end-to-end framework to surpass such paradigms, proving that its cross-view verification RL mechanism is a more effective strategy for ensuring diagnostic reliability than traditional segmentation-based assistance. 
Specifically, E-MRL not only generates a global diagnostic report, but also explicitly localizes a “key evidence slice”—the most informative frame within the 3D CT volume to support its diagnosis. The RL environment extracts this slice for an independent local re-query. If the model describes a tumor in the global report and confirms it in slice verification, our RL mechanism grants an extra reward beyond base reward for lesion attributes. 
This mechanism compels the model to seek visual evidence for local verification in the global reasoning results before generating descriptions, thereby significantly mitigating hallucinations.

Our main contributions are as follows: 
{
\setlength{\parskip}{2pt}
$\bullet$ \textbf{RL Framework:} We propose a novel ``Diagnosis-Localization-Verification'' RL framework, marking the first end-to-end VLM to surpass complex multi-stage segmentation-assisted paradigms and demonstrating that fine-grained evidence-driven reasoning can substantially elevate clinical diagnostic precision.

\begin{sloppypar}
$\bullet$  \textbf{Multi-Granularity Multimodal Reward:} We design a comprehensive reward mechanism that integrates textual diagnostic correctness with visual evidence grounding. By providing dense, multimodal feedback, this approach effectively mitigates hallucinations at both the reasoning and perception levels.
\end{sloppypar}

$\bullet$ \textbf{Cross-view Consistency Alignment:} 
We develop a global-to-local alignment mechanism that synchronizes global diagnostic findings with local slice-level evidence. This ensures rigorous consistency between the generated text and the underlying visual cues.

$\bullet$ \textbf{Substantial Performances and Insights:} Extensive experiments demonstrate that E-MRL achieves synergistic improvements in both diagnostic accuracy and visual evidence localization. Furthermore, our framework and evidence-driven dataset for E-MRL will be open-sourced to foster the development of reliable and trustworthy medical VLMs.

}

\vspace{-2mm}
\section{Related Work}
\vspace{-2mm}
\textbf{Medical Vision-Language Models.} 
Recent Med-LVLMs~\cite{sellergren2025medgemma,li2025eyecaregpt,xie2025heartcare}, including LLaVA-Med~\cite{li2023llava} and HealthGPT~\cite{lin2025healthgpt}, employ supervised fine-tuning (SFT) for aligning vision and text modalities~\cite{hu2024sali}, while M3D~\cite{bai2024m3d}, CT-Chat~\cite{hamamci2026generalist}, and Merlin~\cite{blankemeier2024merlin} extend SFT to 3D images. However, SFT alone can be limited in complex reasoning. Recent works, such as MedVLThinker~\cite{huang2025medvlthinker} and MedVLM-R1~\cite{pan2025medvlm}, incorporate reinforcement learning (RL) with Chain-of-Thought~\cite{litumorchain,fan2026ctrlcot} or multi-agent~\cite{xia2025mmedagent,feng2025doctoragent} strategies to address these challenges. Overall, SFT and RL have become common and effective post-training approaches in Med-LVLMs.

\noindent
 
\textbf{Reinforcement Learning Strategies and Interpretability.}
Most medical RL methods, such as MedVLM-R1~\cite{pan2025medvlm}, Med3D-R1~\cite{lai2026med3d}, and VALOR~\cite{bose2025visual}, lack explicit logical grounding, while MedReason-R1~\cite{li2026medreason} relies on pre-processed suspicious regions as auxiliary inputs. Existing text-based rewards~\cite{lai2026med,pellegrini2025radialog} provide sparse supervision and lead to limited visual grounding and insufficient interpretability. Although prior visual rewards mainly target 2D image localization~\cite{bose2025visual,deria2026medmo}, clinical 3D visual rewards remain absent. To address this gap, we enable interpretable tumor reasoning via 3D evidence-guided multimodal RL.

\vspace{-2mm}
\section{Method}
\vspace{-2mm}
 
\subsection{Overview: The Diagnosis-Localization-Verification Framework}
To address the "black-box" reasoning and visual hallucination issues inherent in medical VLMs, we propose \textbf{Evidence-MRL}, an interpretable reinforcement learning framework with a cross-view consistency verification mechanism.
Our method mimics the radiologists' workflow, formalizing the reasoning process into a closed-loop "Diagnosis-Location-Verification" paradigm:
\textbf{1) Global Diagnosis:} Analyze the entire 3D CT volume and generate a global report with lesion descriptions.
\textbf{2) Evidence Localization:} Identify and output a "Key-Slice" index as visual evidence supporting the diagnosis, ensuring traceability.
\textbf{3) Verification:} Extract the selected key slice, generate a local description, and verify reliability by checking the correspondence of slice indices and lesion attributes.

This decomposition turns end-to-end generation into grounded reasoning, but selecting a discrete slice index is non-differentiable. To enable end-to-end training and visual verification under a unified objective, we formulate the process as a Markov Decision Process (MDP), defined by the tuple $\mathcal{M} = \{\mathcal{S}, \mathcal{A}, \mathcal{P}, \mathcal{R}\}$. 
\textbf{State Space ($\mathcal{S}$)}: At time step $t$, the state $s_t = [I_{ct}, y_{<t}]$ consists of the input 3D CT volume $I_{ct}$ and the history of generated text tokens $y_{<t}$. 
\textbf{Action Space ($\mathcal{A}$)}: To enable interpretability, we define a hybrid action space. An action $a_t$ can be either a text token from the vocabulary (for constructing the report) or a special pointing action $k \in \{1, ..., N\}$, representing the selection of the $k$-th slice as the key visual evidence.
\textbf{Policy ($\mathcal{P} = \pi _{\theta}$)}: We denote a pre-trained VLM (Qwen3-VL) fine-tuned via SFT as the policy model to learn the distribution $\pi_{\theta}(a_t|s_t)$.
\textbf{Reward ($\mathcal{R}$)}: We design a multi-granularity reward, including rewards of Tumor Existence $r_{e}$, Fine-Grained Information $r_{f}$, and Cross-View Consistency $r_c$.
\vspace{-2mm}
\begin{figure}
    \vspace{-4mm}
    \includegraphics[width=0.97\textwidth]{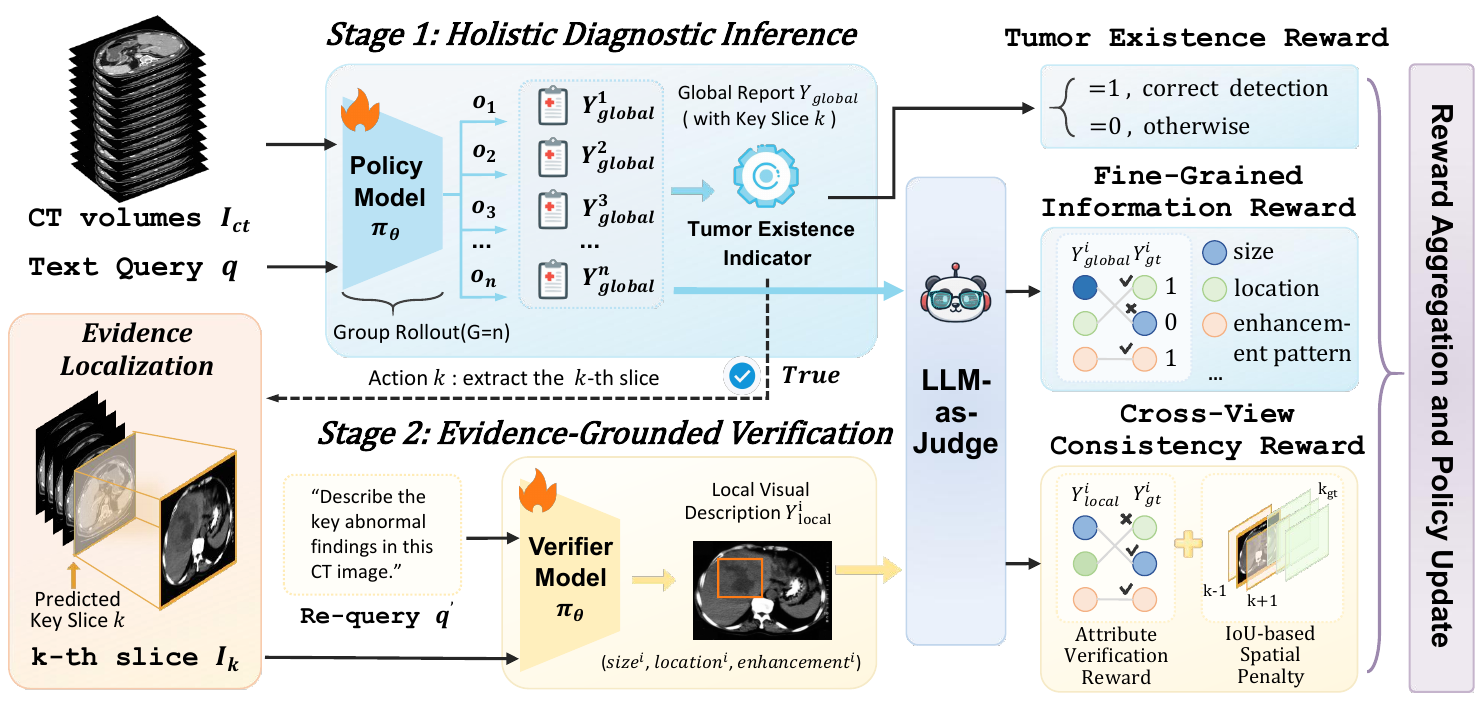}
    \vspace{-0.4cm}
    \caption{The Diagnosis-Localization-Verification framework of Evidence-MRL.} 
     \vspace{-2mm}
    \label{model}
\end{figure}

\vspace{-9mm}
\subsection{Evidence-Driven Decomposed Reasoning}
\vspace{-1mm}
As illustrated in Fig. \ref{model}, our framework operates in two distinct stages.

\noindent
\textbf{Stage 1: Holistic Diagnostic Inference.}
The policy model $\pi_{\theta}$ processes the entire CT sequence to generate a comprehensive diagnostic report $Y_{\text{global}}$ and explicitly predicts a key slice index $k$. The output is structured as follows:
\begin{sloppypar}
\texttt{<findings> f </findings> || <impression> i </impression> || }
\texttt{(opt.) The key lesion is in image <key\_slice> k </key\_slice>, <bbox> [$x_1$, $y_1$, $x_2$, $y_2$] </bbox>}
\end{sloppypar}

\noindent
\textbf{Stage 2: Evidence-Grounded Verification.}
This acts as an environmental feedback step. Once the action $k$ is selected, the environment extracts the $k$-th slice, $I_k$. Subsequently, $I_k$ is fed into a Verifier (a weight-shared copy of the VLM) with a fixed prompt (e.g., ``Describe the abnormality in this image. Should a tumor be identified, please describe its size, location and enhancement pattern.'') to perform a single-image query, yielding a local visual description $Y_{\text{local}}$. This step mimics the radiologist's process of ``double-checking the slice''.

\vspace{-4mm}
\subsection{Multi-Granularity Reward}
\vspace{-1mm}
To drive evidence-based tumor analysis, we formulate a composite reward, deriving textual and visual modalities, with three components: $R_{\text{total}} = \alpha r_e + \beta r_f + \gamma r_{c}$, 
where $\alpha$, $\beta$, $\gamma$ are empirically tuned coefficients, finally set to (0.3, 0.5, 0.2).

\noindent
\textbf{Tumor Existence Reward ($r_{\text{e}}$):} This reward measures binary correctness for tumor existence detection. It is computed against the ground truth labels to ensure baseline clinical safety: $r_e = \mathbb{I} \left[ \mathcal{E}(Y_{\text{global}}) = \mathcal{E}(Y_{\text{gt}}) \right]$,
where $\mathcal{E}(\cdot)$ denotes the tumor existence mapping, yielding 1 if a tumor is present and 0 otherwise, and $\mathbb{I}[\cdot]$ is the indicator function.

\noindent
\textbf{Fine-Grained Information Reward ($r_{\text{f}}$):} 
To promote clinical characterization of tumor lesions, we leverage a powerful LLM (Qwen-235B) to systematically extract structured entities and evaluate three key tumor attributes: size, location, and enhancement pattern. The reward is formulated into a unified metric weighted by clinical significance (coefficients: 0.3, 0.4, and 0.3, respectively):
{
\setlength{\abovedisplayskip}{1pt}   
\setlength{\belowdisplayskip}{1pt}   
\setlength{\abovedisplayshortskip}{0pt}
\setlength{\belowdisplayshortskip}{0pt}
    \begin{equation}
    r_{\text{f}} = \sum_{att \in \Lambda} w_{att} \cdot \mathcal{S}_{att}(Y_{\text{global}}, Y_{\text{gt}})
    \end{equation}
    }
where $\Lambda = \{size, location, enhancement\}$ denotes the set of key attributes and $w_{att}$ reflects the clinical significance of current attribute. Attribute-level LLM scores can be defined as:
{
\setlength{\abovedisplayskip}{1pt}   
\setlength{\belowdisplayskip}{2pt}   
\setlength{\abovedisplayshortskip}{0pt}
\setlength{\belowdisplayshortskip}{0pt}
\begin{equation}
\mathcal{S}_{att}(Y_{\text{global}}, Y_{\text{gt}}) = 
\begin{cases}
1, & \text{if prediction for attribute} \text{ matches gold answer} \\
0, & \text{otherwise}
\end{cases}
\end{equation}
}

\noindent
\textbf{Cross-View Consistency Reward ($r_{\text{c}}$):}
This core contribution is designed to validate the accuracy and interpretability of visual evidence from a local slice perspective. Specifically, we focus on assessing the consistency between the attributes of the largest tumor region in the local slice and those documented in the ground-truth report. It aggregates attribute-level scores from an LLM-as-judge and imposes a spatial penalty based on intersection-over-union (IoU):
{
\setlength{\abovedisplayskip}{1pt}   
\setlength{\belowdisplayskip}{1pt}   
\setlength{\abovedisplayshortskip}{0pt}
\setlength{\belowdisplayshortskip}{0pt}
\begin{equation}
r_c = \sum_{att \in \Lambda} w_{att} \cdot \mathcal{S}_{att}(Y_{\text{local}}, Y_{gt}) - \lambda \cdot \mathcal{L}_{\text{dist}}(k, k_{gt})
\end{equation}
}
where
{
\setlength{\abovedisplayskip}{1pt}   
\setlength{\belowdisplayskip}{1pt}   
\setlength{\abovedisplayshortskip}{0pt}
\setlength{\belowdisplayshortskip}{0pt}
\begin{equation}
\mathcal{L}_{\text{dist}}(k, k_{gt}) = 1 - \text{IoU}([k-1, k+1], [k_{gt}-1, k_{gt}+1])
\end{equation}
}

Here, $k$ and $k_{gt}$ denote the slice number with the maximum tumor cross-sectional area for the prediction and ground truth. To account for potential annotation noise and segmentation inaccuracies, we expand the specific slice index into $[k-1, k+1]$ to calculate IoU. $\lambda$ = 0.3 is a tunable penalty coefficient. 

\vspace{-5mm}
\subsection{Policy Optimization via GRPO}
\vspace{-3mm}
To efficiently optimize the policy model $\pi_{\theta}$ without the overhead of a value network, we employ Group Relative Policy Optimization (GRPO). For each input $I_v$, we sample a group of $G$ outputs $\{o_i\}_{i=1}^G$ from the policy $\pi_{\theta}$, where $o_i = (Y_{\text{global}}^{i}, k^{i}, Y^{i}_{local})$. The advantage $A_i$ is estimated by normalizing rewards within the group:
{
\setlength{\abovedisplayskip}{1pt}   
\setlength{\belowdisplayskip}{5pt}   
\setlength{\abovedisplayshortskip}{0pt}
\setlength{\belowdisplayshortskip}{0pt}
\begin{equation}
    A_i = \frac{R(o_i) - \operatorname{mean}(\{R(o_j)\}_{j=1}^G)}{\operatorname{std}(\{R(o_j)\}_{j=1}^G) + \epsilon}
\end{equation}
}
\vspace{1mm}

The policy is updated by maximizing the surrogate objective with a KL-divergence penalty to maintain linguistic stability:

{
\setlength{\abovedisplayskip}{1pt}   
\setlength{\belowdisplayskip}{1pt}   
\setlength{\abovedisplayshortskip}{0pt}
\setlength{\belowdisplayshortskip}{0pt}
\begin{equation}
    \mathcal{J}_{\text{MRL}}(\theta) = \mathbb{E} \left[ \frac{1}{G} \sum_{i=1}^{G} \left( \min \left( \rho_i A_i, \operatorname{clip}(\rho_i, 1-\varepsilon, 1+\varepsilon) A_i \right) - \beta \mathbb{D}_{\text{KL}}(\pi_{\theta} || \pi_{\text{ref}}) \right) \right]
\end{equation}
}
where $\rho_i = \frac{\pi_{\theta}(o_i | I_v)}{\pi_{\theta_{\text{old}}}(o_i | I_v)}$ is the probability ratio, and $\pi_{\text{ref}}$ is the reference SFT model used to prevent reward hacking.

\section{Experiments}
\vspace{-2mm}
\subsection{Experimental Setup}
\vspace{-2mm}
\noindent
\textbf{Data Preparation.} 
We focus on evidence-driven tumor analysis on a public tumor dataset \textbf{AbdomenAtlas3.0}~\cite{bassi2025radgpt}. 
All samples are cleaned and formatted uniformly and reviewed by experienced radiologists. Based on the tumor segmentation mask, we output key slices as visual evidence in the reports. AbdomenAtlas3.0 contains 9,262 CT images with 1,451 liver, 1,057 pancreas, and 2,049 kidney lesions. We organize each case into a single-organ report, and follows the train-test splitation as \cite{bassi2025radgpt}. Our model is trained on both SFT and RL data derived from the training set and evaluated on test set. To construct a challenging RL dataset, we identify about 2,000 samples with erroneous predictions of tumor attributes or key slice in the SFT outputs for further RL training.

\noindent
\textbf{Implementation Details.}
Qwen-3-VL-8B~\cite{yang2025qwen3} is used as base model of E-MRL. The model first undergoes SFT to acquire essential tumor analysis abilities, during which only the encoder and projection layer parameters are updated. During RL, both the projection layer and LLM parameters are updated and the learning rate is reduced to $5 \times 10^{-6}$ to ensure stable convergence.

\noindent
\textbf{Baselines.}
We compare E-MRL with multiple general and medical models, including two Commercial LVLMs (GPT-4o~\cite{hurst2024gpt} and Gemini 3~\cite{team2023gemini}), three 2D Medical Models (HuluMed~\cite{jiang2025hulu}, Lingshu~\cite{xu2025lingshu} and MedVLM-R1~\cite{pan2025medvlm}), four 3D Medical Models (CT-Chat~\cite{hamamci2026generalist}, CT2Rep~\cite{hamamci2024ct2rep}, Merlin~\cite{blankemeier2024merlin} and M3D~\cite{bai2024m3d}) and RadGPT~\cite{bassi2025radgpt}. 

\noindent
\textbf{Metrics.} Following previous approaches~\cite{bassi2025radgpt}, we use sensitivity and specificity as clinical metrics. To reflect overall clinical ability, we report balanced-accuracy (\textbf{B-Acc}), defined as the mean of sensitivity and specificity. For fine-grained tumor characterization, we assess accuracy for tumor size, location, enhancement pattern, and the hit rate of key slice (\textbf{K-Hit}) using Qwen-235B.
K-Hit evalute the visual evidence reliability by calculating the hit rate based on whether the generated key slice index fall within the range of [$k_{gt}-1$, $k_{gt}+1$].

\vspace{-4mm}
\subsection{Results}
\vspace{-2mm}
\begin{table*}[t]
\caption{Clinical Results (Sen., Spe., and Balanced-Acc.) on AbdomenAtlas3.0 benchmark. \textbf{Bold} denotes the best. \underline{Underline} denotes the second highest scores.}
\vspace{-3mm}
\vspace{-0.7cm}
\label{tab:medical_results}
\begin{center}
  \setlength{\tabcolsep}{2.5pt} 
  \renewcommand{\arraystretch}{1.2}
\resizebox{\linewidth}{!}{
\begin{tabular}{lccccccccccccc}
\toprule
\multirow{3}{*}{\textbf{Model}} 
& \multicolumn{4}{c}{\textbf{Pancreatic (\%)}} 
& \multicolumn{4}{c}{\textbf{Kidney (\%)}} 
& \multicolumn{4}{c}{\textbf{Liver (\%)}} 
& \multirow{3}{*}{\textbf{\begin{tabular}{@{}c@{}}Avg.\\B-Acc.\end{tabular}}} \\
\cmidrule(lr){2-5} \cmidrule(lr){6-9} \cmidrule(lr){10-13}
& \begin{tabular}{@{}c@{}}Sen.\\{\scriptsize($\le$2cm)}\end{tabular} 
& \begin{tabular}{@{}c@{}}Sen.\\{\scriptsize($>$2cm)}\end{tabular} 
& \begin{tabular}{@{}c@{}}Spec.\end{tabular} 
& \begin{tabular}{@{}c@{}}B-Acc.\end{tabular} 
& \begin{tabular}{@{}c@{}}Sen.\\{\scriptsize($\le$2cm)}\end{tabular} 
& \begin{tabular}{@{}c@{}}Sen.\\{\scriptsize($>$2cm)}\end{tabular} 
& \begin{tabular}{@{}c@{}}Spec.\end{tabular} 
& \begin{tabular}{@{}c@{}}B-Acc.\end{tabular} 
& \begin{tabular}{@{}c@{}}Sen.\\{\scriptsize($\le$2cm)}\end{tabular} 
& \begin{tabular}{@{}c@{}}Sen.\\{\scriptsize($>$2cm)}\end{tabular} 
& \begin{tabular}{@{}c@{}}Spec.\end{tabular} 
& \begin{tabular}{@{}c@{}}B-Acc.\end{tabular} 
& \\
\midrule
\rowcolor{gray!5}
\multicolumn{14}{c}{\textit{Commercial VLM Models}} \\
\midrule
GPT-4o-mini & 57.58 & 68.12 & 35.06 & 37.28 & 27.85 & 55.47 & 66.11 & 55.51 & 79.03 & 95.52 & 31.55 & 59.57 & 50.79 \\
Gemini 3 & 96.67 & 98.41 & 9.78 & 54.38 & 94.29 & 100.00 & 8.89 & 53.35 & 87.50 & 98.33 & 13.68 & 53.40 & 53.71 \\
\midrule
\rowcolor{gray!5}
\multicolumn{14}{c}{\textit{2D Medical VLM Models}} \\
\midrule
HuluMed-7B & 96.97 & 97.10 & 6.49 & 51.77 & 73.42 & 74.22 & 34.30 & 54.11 & 46.77 & 77.61 & 60.11 & 61.45 & 55.78 \\
Lingshu-7B & 78.79 & 88.41 & 19.82 & 51.92 & 68.35 & 87.50 & 36.84 & 58.51 & 50.00 & 76.12 & 49.02 & 56.29 & 55.57 \\
MedVLM-R1-2B & 46.88 & 57.35 & 37.82 & 45.19 & 55.70 & 62.20 & 57.89 & 58.80 & 37.78 & 70.37 & 30.89 & 42.80 & 48.93 \\
\midrule
\rowcolor{gray!5}
\multicolumn{14}{c}{\textit{3D Medical VLM Models (\textit{w/} SFT)}} \\
\midrule
CT-Chat & 66.70 & 51.90 & 61.20 & 58.40 & 31.10 & 32.80 & 74.20 & 53.03 & 5.70 & 3.20 & 94.70 & 49.52 & 53.65 \\
CT2Rep & 0.00 & 0.00 & 92.50 & 46.25 & 36.50 & 39.30 & 70.40 & 54.08 & 35.80 & 49.20 & 70.40 & 56.74 & 52.36 \\
Merlin & 33.30 & 51.90 & 71.80 & 59.52 & 28.40 & 45.90 & 86.60 & 61.45 & 30.20 & 41.30 & 95.90 & 66.06 & 62.34 \\
M3D & 8.18 & 18.84 & 51.17 & 32.57 & 11.39 & 16.41 & 53.36 & 33.68 & 22.58 & 26.87 & 61.72 & 43.47 & 36.57 \\
\midrule
\rowcolor{gray!5}
\multicolumn{14}{c}{\textit{Previous SOTA: Segmentation-assisted Report Generation Model}} \\
\midrule
RadGPT & 66.70 & 81.50 & 93.20 & \textbf{85.50} & 54.80 & 93.30 & 51.80 & 62.00 & 39.60 & 96.80 & 64.40 & 67.53 & 71.68 \\
\midrule
\rowcolor{gray!5}
\multicolumn{14}{c}{\textit{Our Models (Backbone: Qwen3-VL-8B)}} \\
\midrule
\rowcolor{violet!5} Zero-shot Inference & 6.06 & 15.94 & 90.27 & 50.85 & 15.19 & 34.38 & 88.02 & 57.53 & 18.03 & 16.18 & 87.48 & 52.27 & 53.55 \\
\rowcolor{violet!5} \begin{tabular}{@{}l@{}}SFT\end{tabular} & 66.67 & 71.01 & 82.74 & 75.88 & 79.75 & 92.97 & 84.57 & 86.24 & 86.57 & 87.10 & 90.31 & 88.58 & 83.57 \\
\rowcolor{violet!5} SFT+GRPO (\textit{w/} $r_e$ )& 60.61 & 78.34 & 85.40 & 77.82 & 77.22 & 92.19 & 89.42 & \underline{87.94} & 83.51 & 89.55 & 95.47 & 91.06 & 85.61 \\
\rowcolor{violet!5} SFT+GRPO (\textit{w/} $r_e$, $r_f$)& 75.76 & 84.06 & 83.82 & \underline{82.04} & 82.28 & 91.41 & 85.08 & 86.50 & 88.52 & 93.03 & 93.78 & \underline{92.32} & \underline{86.95} \\
\rowcolor{blue!5} \begin{tabular}{@{}l@{}}\textbf{SFT+E-MRL}\end{tabular} & 78.79 & 84.06 & 83.21 & \textbf{82.43} & 89.74 & 94.44 & 83.82 & \textbf{88.23} & 90.32 & 92.54 & 94.81 & \textbf{93.14} & \textbf{87.93} \\
\rowcolor{blue!5} $\Delta$ vs SFT model & \textcolor{purple!60!black}{$\uparrow$ 12.12} & \textcolor{purple!60!black}{$\uparrow$ 13.05} & \textcolor{purple!60!black}{$\uparrow$ 0.47} & \textcolor{purple!60!black}{$\uparrow$ 6.55} & \textcolor{purple!60!black}{$\uparrow$ 9.99} & \textcolor{purple!60!black}{$\uparrow$ 1.47} & \textcolor{purple!60!black}{$\downarrow$ 0.75} & \textcolor{purple!60!black}{$\uparrow$ 1.99} & \textcolor{purple!60!black}{$\uparrow$ 3.75} & \textcolor{purple!60!black}{$\uparrow$ 5.44} & \textcolor{purple!60!black}{$\uparrow$ 4.50} & \textcolor{purple!60!black}{$\uparrow$ 4.56} & \textcolor{purple!60!black}{$\uparrow$ 4.36} \\
\bottomrule
\end{tabular}
}
\end{center}
\vspace{-1cm}
\end{table*}

\begin{figure}
    \vspace{-7mm}
    \includegraphics[width=\textwidth]{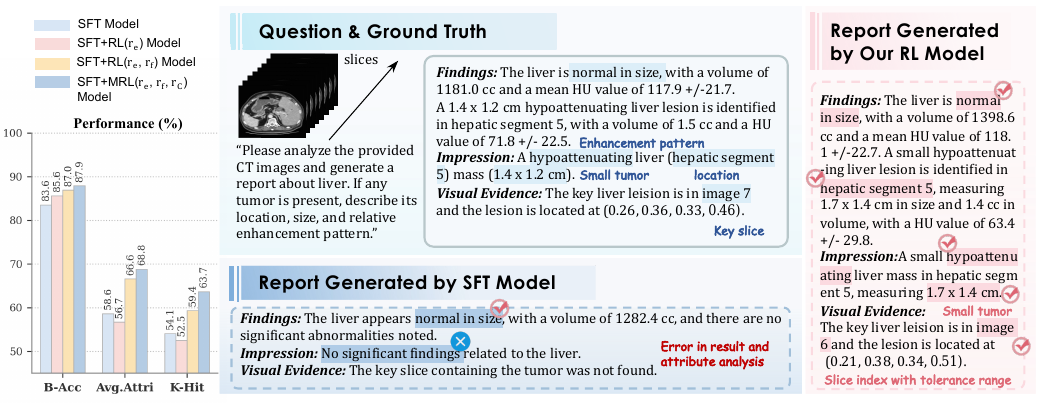}
    \vspace{-0.8cm}
    \caption{(a) \textbf{Ablation} study on three kinds of multi-granularity rewards. (b) A \textbf{case analysis} of performance enhancement through E-MRL built upon the SFT model.} 
    \label{casestudy}
\end{figure}

\begin{table*}[t]
\caption{Lesion Attributes and visual evidence evaluation (size, location, enhancement and K-Hit) on AbdomenAtlas3.0 benchmark. "--" denotes that the model does not have the capability to generate key slice descriptions.}
\vspace{-1cm}
\label{tab:fine_grained_attributes}
\begin{center}
  \setlength{\tabcolsep}{2.5pt} 
  \renewcommand{\arraystretch}{1.2}
\resizebox{\linewidth}{!}{
\begin{tabular}{lccccccccccccc}
\toprule
\multirow{3}{*}{\textbf{Model}} 
& \multicolumn{4}{c}{\textbf{Pancreatic Tumor (\%)}} 
& \multicolumn{4}{c}{\textbf{Kidney Tumor (\%)}} 
& \multicolumn{4}{c}{\textbf{Liver Tumor (\%)}} 
& \multirow{3}{*}{\textbf{Avg.}} \\
\cmidrule(lr){2-5} \cmidrule(lr){6-9} \cmidrule(lr){10-13}
& Size 
& Loc. 
& \begin{tabular}{@{}c@{}}Enhan.\end{tabular} 
& K-Hit 
& Size 
& Loc. 
& \begin{tabular}{@{}c@{}}Enhan.\end{tabular} 
& K-Hit 
& Size 
& Loc. 
& \begin{tabular}{@{}c@{}}Enhan.\end{tabular} 
& K-Hit 
& \\
\midrule
\rowcolor{gray!5}
\multicolumn{14}{c}{\textit{Baseline Models}} \\
\midrule
GPT-4o-mini & 20.59 & 50.00 & 45.10 & 27.45 & 11.11 & 41.06 & 54.59 & 14.49 & 44.96 & 20.93 & 77.52 & 23.26 & 35.92 \\
Gemini 3 & 31.96 & 43.30 & \underline{51.55} & 30.52 & 56.57 & 63.74 & \underline{77.80} & 28.28 & 66.67 & 36.43 & 82.95 & 27.13 & 49.74 \\
HuluMed-7B & 30.39 & 12.75 & 25.49 & 9.49 & 37.20 & 36.71 & 51.69 & 11.32 & 37.98 & 28.68 & 40.31 & 7.85 & 27.49 \\
Lingshu-7B & 26.47 & 20.59 & 26.47 & 19.79 & 28.23 & 11.21 & 15.80 & 13.21 & 38.08 & 8.72 & 12.21 & 10.91 & 19.31 \\
MedVLM-R1-2B & 21.77 & 14.92 & 12.79 & 1.94 & 38.65 & 31.88 & 25.12 & 2.83 & 11.34 & 1.06 & 2.47 & 0.77 & 13.80 \\
M3D & 6.86 & 5.88 & 17.45 & -- & 3.01 & 11.11 & 18.02  & -- & 3.18 & 1.55 & 13.18 & -- & 8.92 \\
\midrule
\rowcolor{gray!5}
\multicolumn{14}{c}{\textit{Our Models (Backbone: Qwen3-VL-8B)}} \\
\midrule
\rowcolor{violet!5} Zero-shot Inference & 4.90 & 15.69 & 32.35 & 1.96 & 9.18 & 32.85 & 46.38 & 2.42 & 10.85 & 3.10 & 37.21 & 1.83 & 16.56 \\
\rowcolor{violet!5} \begin{tabular}{@{}l@{}}SFT\end{tabular} & \underline{50.98} & 53.92 & 49.02 & 48.04 & 64.73 & 73.43 & 70.53 & 56.04 & 65.12 & 25.58 & 74.42 & 58.10 & 57.49 \\
\rowcolor{violet!5} SFT+GRPO (\textit{w/} $r_e$)& 46.27 & 46.08 & 45.20 & 44.12 & 66.18 & 70.46 & 71.50 & 60.80 & 65.12 & 26.36 & 72.87 & 52.64 & 55.63 \\
\rowcolor{violet!5} SFT+GRPO (\textit{w/} $r_e$, $r_f$)& 50.00 & \underline{64.71} & 50.98 & \underline{53.92} & \underline{71.50} & \underline{74.88} & 77.29 & \underline{63.77} & \underline{76.77} & \underline{46.51} & \underline{86.40} & \underline{60.36} & \underline{64.76} \\
\rowcolor{blue!5} \begin{tabular}{@{}l@{}}\textbf{SFT+E-MRL}\end{tabular} & \textbf{51.96} & \textbf{66.67} & \textbf{52.94} & \textbf{54.90} & \textbf{73.91} & \textbf{76.33} & \textbf{79.71} & \textbf{71.01} & \textbf{79.84} & \textbf{49.61} & \textbf{88.37} & \textbf{65.12} & \textbf{67.53} \\
\bottomrule
\end{tabular}
}
\end{center}
\vspace{-1cm}
\end{table*}
\textbf{Diagnosis Performance.}
Table~\ref{tab:medical_results} demonstrates that our approach consistently surpasses end-to-end commercial solutions, as well as specialized 2D and 3D medical VLM models, across three major tumor types in terms of balanced accuracy (B-Acc). The suboptimal tumor detection accuracy (sensitivity and specificity) observed in most baselines underscores the challenges faced by current LVLMs in generating CT reports for clinically significant tasks. To ensure a fair comparison, all competing 3D medical models~\cite{hamamci2026generalist,hamamci2024ct2rep,blankemeier2024merlin,bai2024m3d} were supervised fine-tuned on our curated dataset. Notably, our SFT variant yielded substantial performance gains; we attribute this to our strategy of fine-tuning the image encoder exclusively while treating CT slices as video sequences. This design effectively retains the foundational representation capabilities of Qwen3-VL-8B. Furthermore, ablation studies(see Fig.~\ref{casestudy}a) validate the efficacy of our proposed RL rewards: consistent improvements were observed, with the full ``Evidence-MRL'' configuration (utilizing all three reward functions) achieving state-of-the-art (SOTA) performance of 87.93\% Avg. B-Acc across three types tumor.

Notably, our approach even outperforms the previous SOTA RadGPT~\cite{bassi2025radgpt} by over 16\% in Avg. B-Acc. RadGPT leverages expert tumor segmentation models for fine-grained report generation. Our model is by far the first end-to-end VLM solution to achieve such clinically applicable performance for tumor-related report generation.

\noindent
\textbf{Attribute Reliability and Visual Grounding Abilities.}
Table~\ref{tab:fine_grained_attributes} reveals that both commercial and specialized medical models exhibit limited accuracy in fine-grained tumor grounding. Key attributes, including tumor size, location, relative enhancement, and key slice localization, represent critical characteristics prioritized by radiologists for interpretation, yet remain challenging for automated report generation systems. In contrast, our approach demonstrates consistent performance gains across training stages: from supervised fine-tuning to GRPO optimization incorporating the tumor existence reward ($r_e$), fine-grained information reward ($r_f$), and cross-view consistency reward ($r_c$). These results underscore our model's superiority in generating grounded evidence and enhancing interpretability. Note that our models cannot be compared with RadGPT because we used RadGPT-generated fine-grained reports as references to evaluate the metrics in Table~\ref{tab:fine_grained_attributes}.

\noindent
\textbf{Cases Study.}
Fig.~\ref{casestudy}(b) illustrates the comparison between the SFT model and the E-MRL model in liver CT report generation. The SFT model missed both tumor detection and attribute analysis. In contrast, E-MRL model accurately detected the small tumor and correctly reported its location and enhancement pattern. The key slice was also identified within the tolerance range, showing a high consistency with the ground truth. This case demonstrates the significant advantages of our proposed E-MRL method in improving tumor detection rates and enriching structured details in radiology reports.

\vspace{-1cm}
\section{Conclusion}

In this work, we introduced Evidence-MRL, a multimodal reinforcement learning framework designed to address visual hallucinations in 3D CT tumor analysis. By formulating report generation as a diagnosis-localization-verification process, our method enforces explicit visual grounding through a Cross-View Consistency Reward, ensuring that global diagnostic conclusions are substantiated by local slice evidence. Extensive experiments demonstrate that Evidence-MRL significantly outperforms existing baselines in both diagnostic accuracy and fine-grained attribute grounding. This work sheds light on developing trustworthy and interpretable end-to-end LVLM systems capable of supporting reliable clinical decision-making for tumor analysis.

\section*{Disclosure of Interest}
The authors have no competing interests to declare that are relevant to the content of this article.

\section*{Acknowledgements}
This work has been supported in part by the NSFC (No. 62436007), the China Postdoctoral Science Foundation under Grant Number 2024M752794, the ZJNSF (No. LZ25F020004), the Key Research and Development Projects in Zhejiang Province (No. 2025C01128, 2025C01030, 2025C02156),
Ningbo Yongjiang Talent Introduction Programme (2023A400-G).

%
%
%
\clearpage
\bibliographystyle{splncs04}
\bibliography{ref}

\end{document}